\documentclass[12pt]{article}
\usepackage{amsmath}
\usepackage{amsthm}
\usepackage{amsfonts}
\usepackage[dvips]{epsfig}
\usepackage{graphicx}
\usepackage{amssymb}
\usepackage{cancel}
\usepackage{pstricks} 
\usepackage{lscape}
\usepackage{enumitem}
\usepackage{subfig}

\usepackage{caption}
\usepackage{relsize}

\newcommand{\beq}{\begin{equation}}
\newcommand{\eeq}{\end{equation}}

\makeatletter
\newlength{\apb@width}
\newcommand{\autoparbox}[2][c]{\settowidth{\apb@width}{#2}\parbox[#1]{\apb@width}{#2}}

\newcommand{\Cen}[2]{%
  \ifmeasuring@
    #2%
  \else
    \makebox[\ifcase\expandafter #1\maxcolumn@widths\fi]{$\displaystyle#2$}%
  \fi
}
\makeatother

\begin{document}

\begin{flushright}
{\tt KCL-PH-TH/2018-52}, {\tt CERN-TH/2018-206}  \\
{\tt ACT-01-18, MI-TH-1889} \\
{\tt UMN-TH-3726/18, FTPI-MINN-18/15} \\
\end{flushright}

\vspace{1cm}
\begin{center}
{\bf {\large De Sitter Vacua in No-Scale Supergravity}}
\end{center}
\vspace{0.1cm}


\begin{center}{
{\bf John~Ellis}$^{a}$,
{\bf Balakrishnan~Nagaraj}$^{b}$,
{\bf Dimitri~V.~Nanopoulos}$^{b,c}$ and
{\bf Keith~A.~Olive}$^{d}$
}
\end{center}

\begin{center}
{\em $^a$Theoretical Particle Physics and Cosmology Group, Department of
  Physics, King's~College~London, London WC2R 2LS, United Kingdom;\\
Theoretical Physics Department, CERN, CH-1211 Geneva 23,
  Switzerland}\\[0.2cm]
{\em $^b$George P. and Cynthia W. Mitchell Institute for Fundamental
 Physics and Astronomy, Texas A\&M University, College Station, TX
 77843, USA}\\[0.2cm]
 {\em $^c$Astroparticle Physics Group, Houston Advanced Research Center (HARC),
 \\ Mitchell Campus, Woodlands, TX 77381, USA;\\ 
Academy of Athens, Division of Natural Sciences,
Athens 10679, Greece}\\[0.2cm]
{\em $^d$William I. Fine Theoretical Physics Institute, School of
 Physics and Astronomy, University of Minnesota, Minneapolis, MN 55455,
 USA}
 
 \end{center}

\vspace{1cm}
\centerline{\bf ABSTRACT}
\vspace{0.1cm}

{\small No-scale supergravity is the appropriate general framework for low-energy effective field theories derived from string theory.
The simplest no-scale K\"ahler potential with a single chiral field corresponds to a compactification to flat Minkowski space
with a single volume modulus, but generalizations to single-field no-scale models with de Sitter vacua are also known.
In this paper we generalize these de Sitter constructions to two- and multi-field models of the types occurring in string
compactifications with more than one relevant modulus. We discuss the conditions for stability of the de Sitter solutions
and holomorphy of the superpotential, and give examples whose superpotential contains only integer powers of the
chiral fields.
 }

\vspace{1cm}

\begin{flushleft}
September 2018
\end{flushleft}
\medskip
\noindent

\newpage

\section{Introduction}

If one assumes that $N = 1$ supersymmetry holds down to energies hierarchically smaller than the Planck mass,
low-energy dynamics must be governed by some $N = 1$ supergravity. It is known that the energy density in the
present vacuum is very small compared, e.g., to typical energy scales in the Standard Model. It was therefore
natural to look for $N=1$ supergravity theories that yielded a vanishing cosmological constant without unnatural fine tuning,
and a total scalar potential that is positive definite. The unique K\"ahler potential for such an $N=1$ supergravity model with 
a single chiral superfield $\phi$ (up to canonical field redefinitions) was found in~\cite{CFKN} to be
\begin{equation}
K \; = \; - \, 3 \, \ln \left( \phi + \phi^\dagger \right) \, .
\label{CFKN}
\end{equation}
In~\cite{ELNT} this was dubbed `no-scale supergravity', because the scale of supersymmetry breaking is undetermined
at the tree level, and it was suggested that the scale might be set by perturbative corrections to the effective low-energy
field theory. The single-field model (\ref{CFKN}) was explored in more detail in~\cite{EKN1} (EKN), and the generalization to more
superfields was developed in~\cite{EKN2}~\footnote{For a review of early work on no-scale supergravity, see~\cite{LN}.}. 
It was shown subsequently that no-scale supergravity emerges as the
effective field theory resulting from a supersymmetry-preserving compactification of ten-dimensional
supergravity, used as a proxy for compactification of heterotic string theory \cite{Witten}. 

In recent years interest has grown in the possibility of string solutions in de Sitter space, for at least a couple of practical reasons. 
One is the discovery that the expansion of the Universe is accelerating due to non-vanishing vacuum energy that is
small relative to the energy scale of the Standard Model~\cite{cc}. The other is the growing observational support for
inflationary cosmology~\cite{inf}, according to which the Universe underwent an early epoch of near-exponential quasi-de Sitter
expansion driven by vacuum energy that was large compared with the energy scale of the Standard Model, but still
hierarchically smaller than the Planck scale. At the time of writing there is an ongoing controversy whether string theory 
in fact admits consistent solutions in de Sitter space~\cite{string}. 

If string theory does indeed admit de Sitter solutions and approximate supersymmetry with scales hierarchically smaller than the string scale, 
their low-energy dynamics should be described by some suitable supergravity theory that is capable of incorporating the
breaking of supersymmetry that is intrinsic in de Sitter space.
Since string compactifications yield no-scale supergravity as an effective low-energy field theory, it is natural to investigate
how de Sitter space could be accommodated within no-scale supergravity~\footnote{For other approaches, see~\cite{other}.}. 
This question was studied already in~\cite{EKN1},
and the purpose of this paper is to analyze this question in more detail and generality, extending the previous single-field
analysis of~\cite{EKN1,rs} to no-scale models with multiple superfields that are characteristic of generic string
compactifications. These models may provide a useful guide to the possible forms of effective field
theories describing the low-energy dynamics in de Sitter solutions of string theory, assuming that they exist.

The outline of this paper is as follows. In Section~\ref{1field} we review the original motivation and construction
of no-scale supergravity with a vanishing cosmological constant~\cite{CFKN}, and also review the construction
in~\cite{EKN1,rs} of no-scale supergravity models with non-vanishing vacuum energy. Section~\ref{2fieldmodels} describes
the extensions of these models to no-scale supergravity models with two chiral fields, which have an interesting
geometrical visualization. The de Sitter construction is extended to multiple chiral fields in Section~\ref{Nfieldmodels}.
In each case, we discuss the requirements of stability of the vacuum and holomorphy of the superpotential,
and give examples of models whose superpotentials contain only integer powers of the chiral fields.
Finally, Section~\ref{sec:conx} summarizes our conclusions and presents some thoughts for future work.

\section{ Single-Field Models} \label{1field}

\subsection{No-Scale Supergravity Models}

We recall that the geometry of a $N = 1$ supergravity model is characterized by a K\"ahler potential $K$ that
is a Hermitian function of the complex chiral fields $\phi^i$. The kinetic terms of these fields are
\begin{equation}
K_i^j \; \frac{\partial \phi_i}{\partial x_\mu} \frac{\partial \phi^\dagger_j}{\partial x^\mu} \qquad {\rm where} \qquad K_i^j \; \equiv \; \frac{\partial^2 K}{\partial \phi^i \partial \phi_j^\dagger}
\label{kinetic}
\end{equation}
is the K\"ahler metric. Defining also $K_i \equiv \partial K/\partial \phi^i$ and analogously its complex conjugate $K^i$,
the tree-level effective potential is
\begin{equation}
V \; = \; e^K \left[ K^j K_i^{-1 j} K_i - 3 \right] + \frac{1}{2} D^a D^a \, ,
\label{effpotone}
\end{equation}
where $K_i^{-1 j}$ is the inverse of the K\"ahler metric (\ref{kinetic}) and $\frac{1}{2} D^a D^a$ is the $D$-term
contribution, which is absent for chiral fields that are gauge singlets as we assume here.

In this Section we consider the case of a single chiral field $\phi$, in which case
it is easy to verify that the first term in (\ref{effpotone}) can be written in the form
\begin{equation}
V( \phi) \; = \; 9 \ e^{4 K/3} \ K_{\phi \phi^\dagger}^{-1} \ \partial_\phi \partial_{\phi^\dagger} e^{- K/3} \, .
\label{Vphi}
\end{equation}
It is then clear that the unique form of $K$ with a Minkowski solution, for which $V=0$, is
\begin{equation}
K \; = \; - \, 3 \, \ln \left( f(\phi) + f^\dagger ( \phi^\dagger) \right) \, ,
\label{firstnoscale}
\end{equation}
where $f$ is an arbitrary analytic function. In fact, since physical results are unchanged by canonical field transformations,
one can transform $f(\phi) \to \phi$ and recover the simple form (\ref{CFKN}) of the K\"ahler potential for a no-scale 
supergravity model with a single chiral field.

We note that this K\"ahler potential describes a maximally-symmetric SU(1,1)/U(1) manifold whose K\"ahler curvature
$R_i^j \; \equiv \; \partial_i \partial^j \ln K_i^j$ obeys the simple proportionality relation
\begin{equation}
\frac{R_i^j}{K_i^j}  \; \equiv \; R \; = \; \frac{2}{3} \, ,
\label{twothirds}
\end{equation}
which is characteristic of an Einstein-K\"ahler manifold.

This model was generalized in EKN~\cite{EKN1}, where general solutions for all flat potentials were found. The SU(1,1) invariance in 
Eq. (\ref{firstnoscale}) holds whenever \footnote{We note that in extended SU(N,1) no-scale models~\cite{EKN2} that include $N-1$ matter fields, $y_i$, 
with the K\"ahler potential $K = -3 \alpha \log(\phi + \phi^\dagger - y^i y_i^\dagger/3)$, the K\"ahler curvature becomes $R =  (N+1)/3\alpha$.
Our constructions can be generalized to this case, but such generalizations lie beyond the scope of this paper.}
\begin{equation}
R \; \equiv \; \frac{R_i^j}{K_i^j} \; = \; \frac{2}{3\alpha} \, ,
\label{twonths}
\end{equation}
which corresponds (up to irrelevant field redefinitions) to the extended K\"ahler potential
\begin{equation}
G \; = \; K + \ln W(\phi) + \ln W^\dagger(\phi^\dagger) \, ,
\label{deSitterK}
\end{equation}
where
\begin{equation} \label{1fieldkahler}
    K \; = \; - \, 3 \, \alpha\text{ln}(\phi+\phi^{\dagger}) \, ,
\end{equation}
we assume $\alpha > 0$, and $W(\phi)$ is the superpotential~\footnote{
 Starobinsky-like models with $\alpha \ne 1$ were discussed in \cite{eno7}. Such models were later dubbed $\alpha$-attractors in~\cite{KLR,rs}.}. 
In this case the effective potential is
\begin{equation}
    V \; = \; e^G \left[ G^j K_i^{-1 j} G_i - 3 \right] \, .
\end{equation}
EKN found 3 classes of solutions with a constant scalar potential~\cite{EKN1}, namely
\begin{eqnarray}
1) \qquad W & = & a \qquad {\rm and} \qquad \alpha = 1 \, , \label{ekn1} \\
2) \qquad W & = & a \, \phi^{3 \alpha/2} \, , \label{ekn2} \\
3) \qquad  W & = & a \, \phi^{3\alpha/2} (\phi^{3 \sqrt{\alpha}/2} - \phi^{-3 \sqrt{\alpha}/2})\, .  \label{ekn3}
\end{eqnarray}
Solution 1) corresponds to the $V=0$ Minkowski solution discussed above, whereas solutions
2) and 3) yield potentials that are constant in the real direction, but are in unstable in the imaginary direction.
As we discuss further below, stabilization in the imaginary direction is straightforward and allows these solutions
to be used for realistic models with constant non-zero potentials in the real direction. 
We find that 2) leads to anti-de Sitter solutions with $V = - 3/8^\alpha \cdot a^2$ and 3) leads to de Sitter 
solutions~\footnote{We correct here a typo in the third solution given in \cite{EKN1}.}
with $V = 3 \cdot 2^{2-3\alpha} \cdot a^2$. We note that in the particular case $\alpha = 1$ this reduces to $W = a \, (\phi^3 - 1)$, 
which yields the de Sitter solution discussed in \cite{rs}. This was utilized in \cite{eno9} when making the correspondence
between no-scale supergravity and $R^2$ gravity. 

In the following subsections, we first generalize the Minkowski solution (\ref{ekn1}), and then show that de Sitter solutions can be obtained
as combinations of Minkowski solutions. These aspects of the solutions will subsequently be used to generalize
them to model theories with multiple moduli.

\subsection{Minkowski Solutions}

We consider the $N=1$ no-scale supergravity model with a single complex chiral field $\phi$ described by the K\"ahler potential
given in (\ref{1fieldkahler}) and the superpotential $W(\phi)$ is a monomial of the form
\begin{equation}
    W \; = \; a \, \phi^{n} \, ,
\end{equation}
and we seek the value of $n$ that admits a Minkowski solution with $V = 0$. Defining $\phi \equiv x+iy$, the potential along real field direction $x$ is given by
\begin{equation} \label{1pot}
    V \; = \; \, 2^{-3\alpha} \cdot \left(\frac{(2n-3\alpha)^2}{3\alpha}-3\right)\cdot a^2 \cdot x^{2n-3\alpha} \, .
\end{equation}
We can obtain a Minkowski solution by setting  to zero the term in the brackets:
\begin{equation}
    \frac{(2n-3\alpha)^2}{3\alpha} \; = \; 3.
\end{equation}
Solving the above equation for $n$, we find two solutions~\cite{rs}:
\begin{equation} \label{1fieldsol}
    n_{\pm} \; = \; \frac{3}{2}(\alpha\pm\sqrt{\alpha}) \, .
\end{equation}
We note that $n_- = 0$ for $\alpha = 1$, corresponding to the EKN solution (\ref{ekn1}) listed above. However, we see that in addition to this $n=0$ solution, 
$n=3$ also yields a Minkowski solution with $V=0$ in all directions in field space.

Although such solutions exist for any $\alpha$, for the superpotential to be holomorphic we need $n_- \geq 0$,
which requires $\alpha\geq1$. Clearly, integer solutions for $n$ are obtained whenever $\alpha$ is a perfect square \cite{rs}. 

It is possible to go from one superpotential to another via a K\"ahler transformation:
\begin{equation}
    K \; \longrightarrow \; K+\lambda(\phi)+\lambda^{\dagger}(\phi^{\dagger}),\qquad W \; \longrightarrow \; e^{-\lambda(\phi)}W \, .
\end{equation}
with $\lambda(\phi)=\pm3\sqrt{\alpha}\hspace{0.1cm}\text{ln}\phi$. 
In general, the solutions (\ref{1fieldsol}) can be thought of as corresponding to endpoints of a line segment of length $3\sqrt{\alpha}$
centred at $3\alpha/2$. Though this appears trivial, extensions of this geometric visualization will be useful in the generalizations to multiple fields discussed below.

For $\alpha \ne 1$, the two solutions yield $V=0$ only along the real direction, and
the mass squared of the imaginary component $y$ along the real field direction for $x > 0$ and $y=0$ is given by
\begin{equation}
    m_{y}^2 \; = \; 2^{2-3\alpha} \cdot \frac{(\alpha-1)}{\alpha} \cdot a^2 \cdot x^{\pm3\sqrt{\alpha}},
\end{equation}
 where the $\pm$ in the exponent corresponds to the two solutions $n_{\pm}$. From this it is clear that the Minkowski solutions are stable for $\alpha\geq 1$.


 There are two aspects of the single-field model  that we emphasize here,
 because they generalize in an interesting way to multi-field models. 
 The first is the fact that there are two solutions for $n$ and the second is that, when $\alpha=1$, we get a Minkowski solution
 with a potential that vanishes everywhere.

 \subsection{De Sitter Solutions}

As was shown in EKN,  de Sitter solutions can be found with the K\"ahler potential (\ref{1fieldkahler}) and a superpotential of the form (\ref{ekn3}),
which may be written as
\begin{equation} \label{1dSsuper}
    W \; = \; a \, (\phi^{n_{-}}-\phi^{n_{+}}) \, ,
\end{equation}
where $n_{\pm}$ are given in (\ref{1fieldsol}). In this case the potential along the real field direction $y=0$ is
\begin{equation} \label{1dSpot}
    V \; = \; 3 \cdot 2^{2-3\alpha} \cdot a^2 \, .
\end{equation}
Thus, the de Sitter solution is obtained by taking the difference of the two ``endpoint" solutions mentioned above. 

\begin{figure}[h!]
\centering
\includegraphics[scale=0.9]{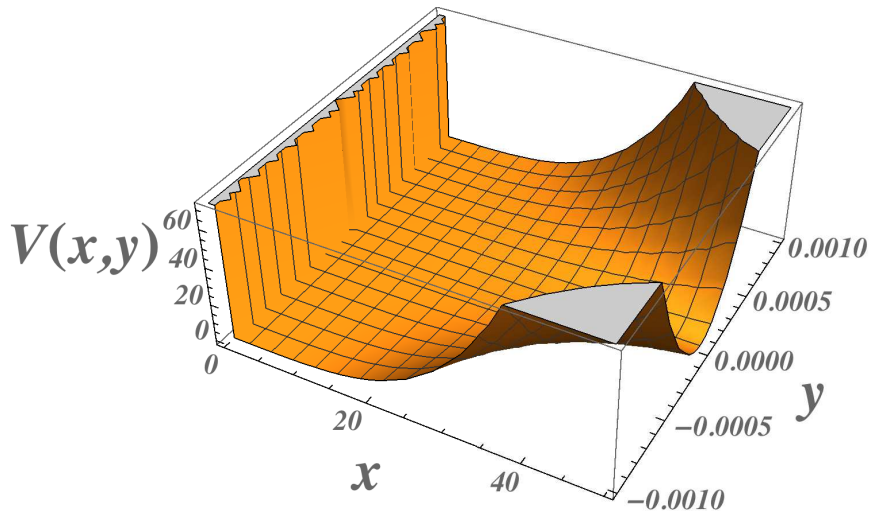}
\caption{\it The potential $V(x,y)$ for $a = b = \alpha=1$ in no-scale supergravity with the stabilized K\"ahler potential
(\ref{stabilized}) and the superpotential (\ref{1dSsuper}). }
\label{deSitter1}
\end{figure}

Unfortunately, this de Sitter solution is not stable for finite $\alpha$. However, this can be remedied by 
deforming the K\"ahler potential to the following form \cite{EKN3,eno7}:
\begin{equation}
    K \; = \; - \, 3 \, \alpha \, \text{ln}(\phi+\phi^{\dagger}+ b (\phi-\phi^{\dagger})^4): \; b \; > \; 0 \; .
    \label{stabilized}
\end{equation}
The addition of the quartic stabilization term does not modify the potential in the real direction,
which is still given by (\ref{1dSpot}).
However, the squared mass of the imaginary component $y$ is now given by
\begin{equation}
    m_y^2 \; = \; \frac{2^{2-3\alpha}}{\alpha} \cdot a^2 \cdot x^{-3\sqrt{\alpha}} \cdot \left(\alpha(x^{3\sqrt{\alpha}}-1)^2-(1-96 b x^3)(x^{3\sqrt{\alpha}}+1)^2\right) \, .
\end{equation}
The stability requirement $m_y^2\geq 0$ is achieved when $\alpha\geq 1$. 
In Fig.~\ref{deSitter1} we plot the stabilized potential for $a = b = \alpha=1$, and we see that
the potential is completely flat along the line $y=0$ and is stable for all values of $x>0$.

\section{Two-Field Models} \label{2fieldmodels}

Several of the features of the single-field model that we discussed in Section~\ref{1field} 
generalize in an interesting geometrical way to models with 
$N > 1$ fields. We illustrate this first by considering in this Section the simplest generalization, i.e., two-field models.

\subsection{Minkowski Solutions}

We consider the following K\"ahler potential with two complex chiral fields:
\begin{equation} \label{2Kahler}
    K \; = \; - \, 3 \, \sum_{i=1}^{2} \, \alpha_i\text{ln}(\phi_i+\phi^{\dagger}_i): \; \alpha_i \; > \; 0 \, .
\end{equation}
with the following ansatz for the superpotential
\begin{equation}
    W= \, a \, \prod_{i=1}^{2}\phi_i^{n_i} \, .
\end{equation}
Denoting the real and imaginary parts by $\phi_i=x_i+iy_i$, we find that the potential along the real field directions $y_i=0$ is given by
\begin{equation}
    V= \,  \left(\sum_{i=1}^2\frac{(2n_i-3\alpha_i)^2}{3\alpha_i}-3\right)\cdot a^2 \cdot \left(\prod_i^22^{-3\alpha_i}x_i^{2n_i-3\alpha_i}\right) \, .
\end{equation}
We see immediately that by setting
\begin{equation} \label{2ellipse}
    \sum_{i=1}^2\frac{(2n_i-3\alpha_i)^2}{3\alpha_i} \; = \; 3
\end{equation}
we obtain a Minkowski solution,

We observe that (\ref{2ellipse}) describes an ellipse in the $(n_1, n_2)$ plane centred at $(3\alpha_1/2,3\alpha_2/2)$. 
All choices of $(n_1, n_2)$ lying on this ellipse yield a Minkowski solution.
In this way, the line segment centred at $3\alpha/2$ in the single-field model that yielded
Minkowski endpoints is generalized, and we 
obtain a one-dimensional continuum subspace of Minkowski solutions. 
We can conveniently parametrize the solutions for $n_i$ in (\ref{2ellipse}) as 
the points on the ellipse corresponding to unit vectors $\vec{r}=(r_1,r_2)$ with $r_1^2+r_2^2=1$:
\begin{equation} \label{2fieldsol}
    n_{i\pm} \; = \; \frac{3}{2}\left(\alpha_i\pm\frac{r_i}{\sqrt{\sum_{j=1}^2\frac{r_j^2}{\alpha_j}}}\right) \, , \qquad i=1,2 \, .
\end{equation}
The unit vector $\vec{r}$ should be located starting at the centre of the ellipse, and defines a direction on its circumference. 
The operation $\vec{r}\rightarrow-\vec{r}$ in equation (\ref{2fieldsol}) takes a point on the ellipse to its antipodal point,
an observation we use later to construct de Sitter solutions.  
We note also that holomorphy requires both $n_1, n_2 \geq 0$, i.e.
\begin{equation} \label{2holo}
    \alpha_i+\frac{r_i}{\sqrt{\sum_{j=1}^2\frac{r_j^2}{\alpha_j}}} \; \geq \; 0 \, , \qquad i=1,2 \, .
\end{equation}
As in the case of the  
single-field model, we can move from one point on the ellipse to another point via a K\"ahler transformation. 
This is possible because the superpotential is just a monomial.

Integer solutions for the values of $n_i$ are also possible in the two-field case.
The full set of solutions in the single-field case are valid for ${n_1}_\pm$ when ${n_2}_+ = {n_2}_- $
(and similarly when $1 \leftrightarrow 2$). More generally, solutions can be found by writing 
\begin{equation}
(n_{1+}-n_{1-})^2 \; = \; \lambda_1(n_{1+}+n_{1-}) \qquad {\rm and} \qquad
(n_{2+}-n_{2-})^2 \; = \; \lambda_2(n_{2+}+n_{2-}) \, ,
\label{twointeger}
\end{equation}
with $\lambda_i$ is non-negative and $\lambda_1 + \lambda_2 = 3$. 
As one example out of an infinite number of solutions,
choosing $\lambda_1 = 1$ and $\lambda_2 = 2$ gives $({n_1}_+,{n_1}_-) = (3,1)$ and 
$({n_2}_+,{n_2}_-) = (6,2)$.

In general, points around the ellipse yield potentials that are flat only in the real direction and,
as in the single-field model, may not be stable in the imaginary directions. 
The masses of the imaginary component fields $y_1, y_2$ are given by
\begin{equation}
    m_{y_i}^2 \; = \;  \,  \frac{2^{2-3(\alpha_1+\alpha_2)}}{\alpha_i^2} \cdot \left(\alpha_i^2-\frac{r_i^2}{\left(\sum_{j=1}^2\frac{r_j^2}{\alpha_j}\right)}\right) \cdot a^2 \cdot x_1^{2n_{1}-3\alpha_1}x_2^{2n_{2}-3\alpha_2} \, , \qquad i=1,2 \, .
\end{equation}
The stability requirement $m_{y_i}^2\geq 0$ for $x_i > 0$ implies
\begin{equation} \label{2stability}
    \alpha_i^2-\frac{r_i^2}{\left(\sum_{j=1}^2\frac{r_j^2}{\alpha_j}\right)} \; \geq \; 0 \, , \qquad i=1,2 \, .
\end{equation}
It is easy to see that if the stability conditions are satisfied then the holomorphy conditions (\ref{2holo}) are satisfied.  Since the left hand side of (\ref{2stability}) is proportional to 
${n_i}_+ {n_i}_-$, points on the ellipse that give stable Minkowski solution are those that are holomorphic so long as their antipodal points are also holomorphic.

\begin{figure}[!ht]
  \centering
{\includegraphics[width=0.5
 \textwidth]{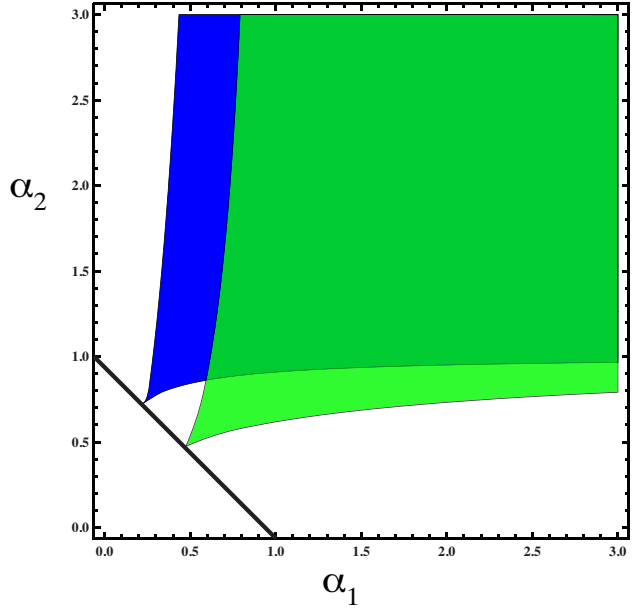}
  \caption{\it The shaded regions are the allowed values of $\alpha_1,\alpha_2$ for 
  the illustrative choices $\vec{r}=(1/\sqrt{2},1/\sqrt{2})$ (green) and  $\vec{r}=(1/\sqrt{10},3/\sqrt{10})$ (blue). 
  There are kinks located at $(\alpha_1,\alpha_2)=(1/2,1/2)$ and $(\alpha_1,\alpha_2)=(1/4,3/4)$
  for the two choices of unit vectors.  The black line is $\alpha_1+\alpha_2=1$.}
  \label{f1}}
\end{figure}

However, given a choice of unit vector, $\vec{r}$, this condition is not satisfied for all choices  of $\alpha_i$. 
We show in Fig.~\ref{f1} the allowed domain in the $(\alpha_1, \alpha_2)$ plane 
for which the stability conditions (\ref{2stability}) (and hence also the holomorphy conditions 
(\ref{2holo})) are satisfied, for two illustrative choices of the unit vector $\vec{r}$. 
The allowed region for $\vec{r}=(1/\sqrt{2},1/\sqrt{2})$ is shaded green and behind it (shaded blue)
is the allowed region when $\vec{r}=(1/\sqrt{10},3/\sqrt{10})$.  For both choices of $\vec{r}$, 
there is a kink in the allowed domain where it meets
the line given by $\alpha_1+\alpha_2=1$. At the kink, for all choices of $\vec{r}$, the potential is completely flat and 
vanishes in all directions in field space. 
The position of the kink can be calculated by solving the stability condition along this line:
\beq
\alpha_1 \; = \; \frac{r_1^2 - \sqrt{r_1^2-r_1^4}}{2r_1^2 - 1} \, .
\label{a1}
\eeq
For the two examples shown in Fig.~\ref{f1},  $r_1 = 1/\sqrt{2}$ implies $\alpha_ 1 = 1/2$ at the kink, 
and $r_1 = 1/\sqrt{10}$ implies $\alpha_ 1 = 1/4$.
In fact, because of the sign ambiguity, there are four unit vectors for each solution, 
corresponding to the ambiguous signs of $r_1$ and $r_2$. 

\begin{figure}[!ht]
  \centering
{\includegraphics[width=0.5
 \textwidth]{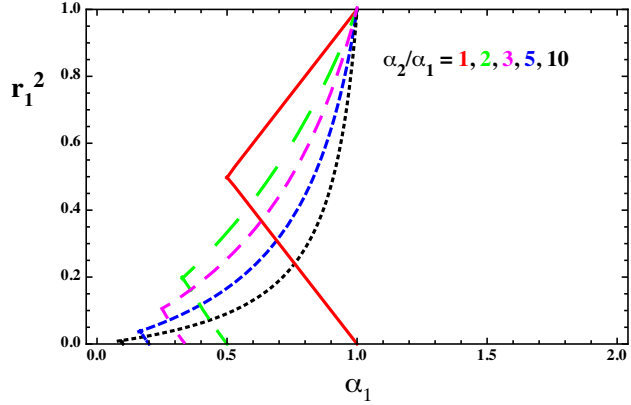}
  \caption{\it The allowed values of $\alpha_1,r_1^2$ for fixed ratios of $\alpha_2/\alpha_1 = 1, 2, 3, 5, 10$. 
  The two sets of curves are derived from the 
  two constraint equations in (\ref{2stability}). The stability inequality is satisfied for points with $\alpha_1$ to the right of both curves
  of the same colour (red, green, purple, blue and black for increasing $\alpha_2/\alpha_1$). The point at which the two curves meet corresponds to 
  the kink that appears in Fig~ \ref{f1} when $\alpha_1 + \alpha_2 = 1$. }
  \label{f2}}
\end{figure}

Another projection of the domain of stability is shown in Fig.~\ref{f2}, which displays the allowed regions in the
($\alpha_1,r_1^2$) plane for the fixed values $\alpha_2/\alpha_1 = 1, 2, 3, 5, 10$, as illustrated by the  curves, respectively.
Each pair of curves (red, green, purple, blue and black for increasing $\alpha_2/\alpha_1$) corresponds to the two equalities in (\ref{2stability}), 
and the positivity inequalities are satisfied to the right of each pair of lines for a given 
value of $\alpha_2/\alpha_1$.  For example, when $\alpha_2/\alpha_1 = 1$ (shown by the solid red curves), 
all values of $r_1^2$ are allowed if $\alpha_1 \ge 1$,
while no values are allowed when $\alpha_1 < 1/2$. The point where the curves meet 
corresponds to the kink when $\alpha_1 = \alpha_2 = 1/2$ and $r_1^2 = 1/2$ that
was seen in Fig. \ref{f1} where the green shaded region touches the black line. 
When $\alpha_2/\alpha_1 = 3$ (shown by medium dashed purple curves), the kink occurs when
these two curves meet at $\alpha_1 = 1/4$ and $r_1^2= 1/10$.

\begin{figure}[ht]
\centering
\includegraphics[scale=0.5]{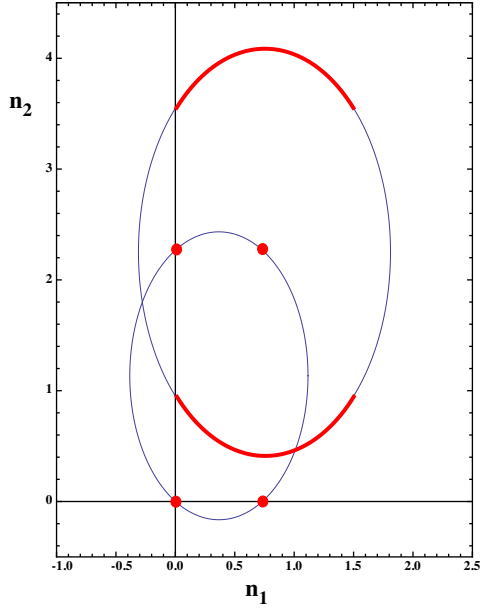}
\caption{\it Minkowski solutions for $\alpha_1=1/4$, $\alpha_2=3/4$ (lower ellipse) and
$\alpha_1=1/2$, $\alpha_2=3/2$ (upper ellipse). In the former case only the four red points 
corresponding to $\vec{r}=(\pm1/\sqrt{10},\pm3/\sqrt{10})$ are allowed, whereas in the
latter case the red arc segments correspond to allowed solutions.}
\label{dS2}
\end{figure}

The lower ellipse (\ref{2ellipse}) in the $(n_1,n_2)$ plane shown in Fig.~\ref{dS2} corresponds to this second example. 
As this corresponds to the position of the kink, only a single value of $r_1^2 = 1/10$ is allowed. 
The four red spots in the figure correspond to the four different vectors $\vec{r}=(\pm1/\sqrt{10},\pm3/\sqrt{10})$.
These four unit vectors correspond to four different superpotentials via the relation (\ref{2fieldsol}), which give
$(n_1,n_2) = (3/4,9/4), (3/4,0), (0,9/4), (0,0)$. When $(\alpha_1,\alpha_2) = (1/4,3/4)$, 
each of the four superpotentials defined by the pair $n_i$
yields a true Minkowski solution. However, because we are at the kink, there are no other stable solutions.

Choosing a larger value of $\alpha_1$ while keeping $\alpha_2/\alpha_1$ fixed
would increase the allowed range in $r_1^2$ (as seen in Fig.~\ref{f2})
and would allow a continuum of stable Minkowski solutions along the real direction in field space. 
This is seen in the upper ellipse in Fig.~\ref{dS2}, where we have chosen $\alpha_1=1/2$ and $\alpha_2=3/2$.
In this case, the stability constraint, which can be read off Fig.~\ref{f2} for $\alpha_2/\alpha_1 = 3$ at the chosen value of $\alpha_1$, yields
$r_1 < 1/2$. Unit vectors with $r_1 < 1/2$ correspond to arcs along the upper ellipse in Fig.~\ref{dS2}.
These are further shortened by the holomorphy requirement that $n_i \ge 0$, 
and the resulting allowed solutions are shown by the red arc segments in the upper ellipse.\\
 
To summarize this discussion of Minkowski solutions in the two-field case: \\
1) For any generic unit vector $\vec{r}$, there is always a kink in the boundary of the 
allowed values of $(\alpha_1, \alpha_2)$ as shown in Fig.~\ref{f1}, and these kink solutions 
always satisfy $\alpha_1+\alpha_2=1$ with $\alpha_1$ given by
(\ref{a1}). The kink solutions give a vanishing potential $V=0$ in all directions in field space.
\\
2) For any pair $(\alpha_1,\alpha_2)$ satisfying $\alpha_1+\alpha_2=1$, 
there are four unit vectors that are determined by inverting (\ref{a1}), namely 
\beq
r_1 = \pm \frac{\alpha_1}{\sqrt{1-2\alpha_1+2\alpha_1^2}} \, .
\eeq
The four values of the $n_i$ that correspond to these choices are $(n_1, n_2) = (0,0),
(3 \alpha_1, 0), (0, 3\alpha_2), \\ (3\alpha_1, 3\alpha_2)$. \\
3) For $\alpha_1 + \alpha_2 > 1$, a continuum of stable Minkowski solutions exist and,
when $\alpha_1 \ge 1$ with $\alpha_2/\alpha_1 \ge 1$,
the entire ellipse (that is, all unit vectors $\vec{r}$) yield stable Minkowski solutions in the real directions of field space.\\
4) The holomorphy conditions (\ref{2holo}) are satisfied automatically if the stability conditions (\ref{2stability}) are satisfied.\\
5) There is an infinite set of Minkowski solutions with positive integer powers of the fields in the superpotential.

\subsection{De Sitter Solutions}

We recall that in the single-field model we were able to construct a de Sitter solution 
by combining the two superpotentials corresponding to Minkowski solutions that can be
visualized as opposite ends of a line segment. In the two-field model, 
we have a continuum of superpotentials that give Minkowski solutions, which are described by an ellipse (\ref{2ellipse}). 
In this case it is possible to  to construct new de Sitter solutions by combining superpotentials corresponding to 
antipodal points on the ellipse (\ref{2ellipse}).
For example, consider the following combined superpotential:
\begin{equation}
    W= \, a \, \left(\phi_1^{n_{1+}}\phi_2^{n_{2+}}-\phi_1^{n_{1-}}\phi_2^{n_{2-}}\right) \, .
\end{equation}
It is easy to see that the scalar potential in the real field direction is a de Sitter solution:
\begin{equation} \label{2pot}
    V \; = \; 3 \cdot 2^{2-3\alpha_1-3\alpha_2} \cdot a^2
\end{equation}
in this case.

For the example described by the lower ellipse in Fig.~\ref{dS2}, one example of a de Sitter solution
is found by taking antipodal points corresponding to the red spots.
When  $\vec{r}=(1/\sqrt{10},3/\sqrt{10})$, we have $W = a (\phi_1^{3/4} \phi_2^{9/4} - 1)$,
which is the unique solution with a holomorphic superpotential that results in a flat de Sitter potential in the real direction.
However, as we discuss further below, this solution is actually not stable.

\begin{figure}[ht]
\centering
\includegraphics[scale=0.5]{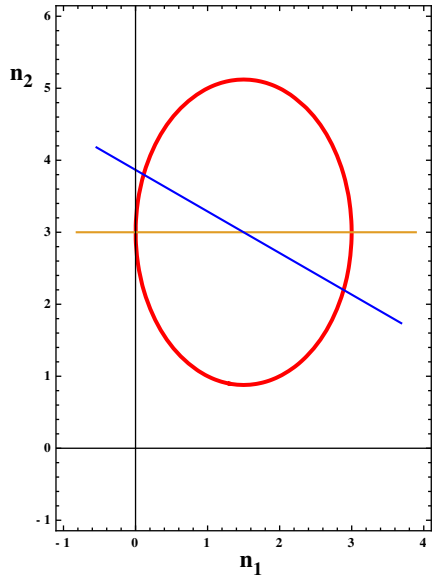}
\caption{\it The Minkowski solutions for $\alpha_1=1$ and $\alpha_2=2$ are described by an ellipse in $(n_1,n_2)$ space. 
Lines passing through the center of the ellipse connect antipodal points, as illustrated with two examples.}
\label{fff1}
\end{figure}

As an alternative example, we consider a two-field model with $\alpha_1=1$ and $\alpha_2=2$.
The Minkowski solutions in this case are described by the ellipse (\ref{2ellipse}) in $(n_1,n_2)$ space
shown in Fig.~\ref{fff1}, whose centre is at $(3/2,3)$. In this case, the entire ellipse can be used to 
construct de Sitter solutions, as all possible unit vectors $\vec{r}$ are allowed since $\alpha_1>1$ (see Fig.~\ref{f2}).
As in the previous example, we can use antipodal points to construct de Sitter solutions, as illustrated in Fig.~\ref{fff1}.
 One such pair of antipodal points is $(3,3),(3,0)$, corresponding to $\vec{r}=(1,0)$, indicated by
 the horizontal orange line in Fig.~\ref{fff1}. The corresponding
 superpotential is
 \begin{equation}
    W \; = \; a^2 \, \left(\phi_1^3\phi_2^3-\phi_2^3 \right) \, ,
\end{equation}
so that the fields appear in the superpotential with positive integer powers. 
This example yields a de Sitter potential with the potential value
\begin{equation}
    V \; = \; 3 \cdot 2^{-7} \cdot a^2.
    \label{ds3}
\end{equation}
along the real field directions. A continuum of de Sitter solutions for real field values 
are possible for different choices of $\vec{r}$, e.g., the choice indicated in Fig.~\ref{fff1} by the blue line, all with the potential given by Eq. (\ref{ds3}).

\subsection{Stability Analysis}

As in the single-field case, the de Sitter solutions of the two-field model require modification in order to be stable.
Stable solutions can easily be found by deforming the K\"ahler potential to include stabilizing quartic terms:
\begin{equation}
    K \; = \; - \, 3 \, \sum_{i=1}^{2} \, \alpha_i\text{ln}(\phi_i+\phi^{\dagger}_i+ b_i (\phi_i-\phi_i^{\dagger})^4): \; b_i \; > \; 0 \, .
\end{equation}
With this modification the potential along real field directions is still given by equation (\ref{2pot}). 
To prove the stability of the two-field de Sitter solution with the quartic modification of the K\"ahler potential, 
we calculate the Hessian matrix $\partial^2 V / \partial y_i \partial y_j: i, j = 1, 2$
along the real field directions, and demand that it be positive semi-definite. The Hessian matrix along the real field directions is of the form
\begin{equation} \label{HMform}
a^2\left(\frac{3.2^{1-3\alpha_1-3\alpha_2}}{\alpha_2r_1^2+\alpha_1r_2^2}\right)
\begin{bmatrix}
    x_1^{-2}A_1 & x_1^{-1}x_2^{-1}B \\
    x_1^{-1}x_2^{-1}B & x_2^{-2}A_2 
\end{bmatrix} \, ,
\end{equation}
where
\begin{equation} 
    A_1\; = \; w^{-1}\left(\alpha_1^2r_2^2(1+4w+w^2)+\alpha_1\alpha_2r_1^2(1-w)^2+\alpha_2r_1^2(96b_1x_1^3-1)(1+w)^2\right) \, ,
\end{equation}
\begin{equation} 
    A_2 \; = \; w^{-1}\left(\alpha_2^2r_1^2(1+4w+w^2)+\alpha_1\alpha_2r_2^2(1-w)^2+\alpha_1r_2^2(96b_2x_2^3-1)(1+w)^2\right) \, ,
\end{equation}
\begin{equation}
    B \; = \; -6\alpha_1\alpha_2r_1r_2 \, ,
\end{equation}
we have defined
\begin{equation} \label{wdef}
    w \; \equiv \; x_1^{\frac{3r_1}{\sqrt{\frac{r_1^2}{\alpha_1}+\frac{r_2^2}{\alpha_2}}}}x_2^{\frac{3r_2}{\sqrt{\frac{r_1^2}{\alpha_1}+\frac{r_2^2}{\alpha_2}}}} \, ,
\end{equation}
and the Hessian matrix is positive semi-definite if the condition
\begin{equation}
{\cal H} \; \equiv \; A_1 \, A_2 \; \geq \; B^2
\label{H+}
\end{equation}
is satisfied.

The stability condition (\ref{H+})  for generic $\alpha_1$, $\alpha_2$, $b_1$, $b_2$ and $\Vec{r}$ is
\begin{equation} \label{Generic}
    \begin{split}
        &\left(\alpha_1^2r_2^2(1+4w+w^2)+\alpha_1\alpha_2r_1^2(1-w)^2+\alpha_2r_1^2(96b_1x_1^3-1)(1+w)^2\right)\\
        \times &\left(\alpha_2^2r_1^2(1+4w+w^2)+\alpha_1\alpha_2r_2^2(1-w)^2+\alpha_1r_2^2\left(96b_2\frac{w^{\frac{1}{r_2}\sqrt{\frac{r_1^2}{\alpha_1}+\frac{r_2^2}{\alpha_2}}}}{x_1^{(3r_1/r_2)}}-1\right)(1+w)^2\right)\\
        &-36\alpha_1^2\alpha_2^2r_1^2r_2^2w^2 \; \geq \; 0 \, .
    \end{split}
\end{equation}
A general stability analysis is intractable, so we have considered the simplified case: 
$\alpha_1=\alpha_2 \equiv \alpha$ and $\Vec{r}=(1/\sqrt{2},1/\sqrt{2})$,
for which the positivity condition (\ref{H+}) becomes
 \begin{equation} \label{Scond1}
     \begin{split} &\left(2\alpha(1+w+w^2)+(96b_1x_1^3-1)(1+w)^2\right)\\
     & \times \left(2\alpha(1+w+w^2)+(96b_2x_2^3-1)(1+w)^2\right) \; \geq \; 36\alpha^2w^2 \, .
     \end{split}
\end{equation}
Eliminating $x_2$ in favour of $x_1$ and $w$ via equation (\ref{wdef}), this inequality becomes 
\begin{equation} \label{Scond2}
    \begin{split} &\left(2\alpha(1+w+w^2)+(96b_1x_1^3-1)(1+w)^2\right)\\
     & \times \left(2\alpha(1+w+w^2)+\left(96b_2\frac{w^{\sqrt{2/\alpha}}}{x_1^3}-1\right)(1+w)^2\right)-36\alpha^2w^2 \; \geq \;0 \, .
     \end{split}
\end{equation}
We note that $(96b_1x_1^3-1)$ dominates for $x_1\gg 1$ and $\left(96b_2\frac{w^{\sqrt{2/\alpha}}}{x_1^3}-1\right)$ dominates for $x_1\ll1$, 
implying that there is an extremum for some intermediate value of $x_1$. This occurs at $x_1=\left(b_2/b_1\right)^{1/6}w^{1/(3\sqrt{2\alpha})}$,
and is a global extremum. Whether it is a maximum or a minimum depends on the sign of $2\alpha(1+w+w^2)-(1+w)^2$, and it is non-negative for
\begin{equation}
   \alpha \geq \frac{2}{3} \, .
\end{equation}
This is a necessary condition for the inequality (\ref{Scond2}) to be satisfied.
We have not explored the full range of possible values of $b_1$ and $b_2$ when $\alpha_1 = \alpha_2 = \alpha$, 
but have checked that the stability condition (\ref{Scond2}) is always satisfied
if $b_1=b_2=1$ and $\alpha \geq 2/3$, irrespective of the value of $w$.
We have also found that when $\alpha_1 \ne \alpha_2$ the sum $\alpha_1 + \alpha_2 \ge 4/3$.

We have also considered the case $\Vec{r}=(0,1)$ with $b_1=b_2=1$. The inequality (\ref{Generic}) reduces
in this case to
 \begin{equation}
    \alpha_2(1-w)^2+(1+w)^2(96w^{1/\sqrt{\alpha_2}}-1) \; \geq \; 0 \, ,
 \end{equation}
which is always satisfied for $\alpha_2 \; \geq \; 1$. It is easy to check that the same is true for the case $\Vec{r}=(1,0)$.
Based on these cases and the previous example with $\Vec{r}=(1/\sqrt{2},1/\sqrt{2})$,
we expect that there are generic stable solutions for a range of $\Vec{r}$ in the first and third quadrants
where $r_1/r_2 > 0$.
However, the situation is different when $r_1/r_2<0$. We find that the inequality (\ref{H+}) cannot be satisfied
for $\Vec{r}=(-1/\sqrt{2},1/\sqrt{2})$ and $b_1=b_2=1$, so there are no stable de Sitter solutions,
and we expect the same to be the case for other choices of $\Vec{r}$ in the second or fourth quadrant.


In summary, we have established the existence of stable de Sitter solutions only when $\Vec{r}$ is in either first or third quadrant.

\section{N-field models} \label{Nfieldmodels}

Finally, we generalize the above set of examples to models with multiple fields $N > 2$.

\subsection{Minkowski Solutions}

\begin{figure}[ht]
\centering
\includegraphics[scale=0.6]{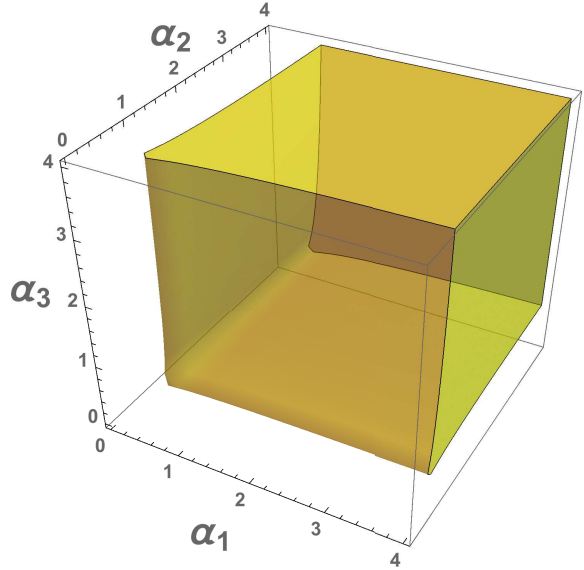}
\caption{\it Allowed values of $\alpha_1,\alpha_2,\alpha_3$ for a three-field model with $\vec{r}=(1/\sqrt{3},1/\sqrt{3},1/\sqrt{3})$.}
\label{q44}
\end{figure}

The natural generalization of the K\"ahler potential in (\ref{2Kahler}) is simply a sum of $N$ similar terms:
\begin{equation}
    K \; = \; - \, 3 \, \sum_{i=1}^{N} \, \alpha_i\text{ln}(\phi_i+\phi^{\dagger}_i) \, .
\end{equation}
Similarly, we adopt the following ansatz for the superpotential:
\begin{equation}
    W \; = \; a \, \prod_{i=1}^{N}\phi_i^{n_i} \, ,
\end{equation}
in which case the potential along the real field directions $x_i$ is 
\begin{equation}
    V\; = \; \left(\sum_{i=1}^N\frac{(2n_i-3\alpha_i)^2}{3\alpha_i}-3\right) \cdot a^2 \cdot \left(\prod_i^N2^{-3\alpha_i}x_i^{2n_i-3\alpha_i}\right) \, .
\end{equation}
We can obtain Minkowski solutions along the real field directions by setting
\begin{equation} \label{Nellipse}
    \sum_{i=1}^N\frac{(2n_i-3\alpha_i)^2}{3\alpha_i} \; = \; 3 \, ,
\end{equation}
which describes an ellipsoid in $(n_1,...,n_N)$ space whose centre is at $(3\alpha_1/2,...,3\alpha_N/2)$. 
Once again we find a continuum of Minkowski solutions. The points on the ellipsoid can be parametrized 
conveniently using an $N$-dimensional unit vector $\vec{r}$:
\begin{equation}
    n_{i} \; = \; \frac{3}{2}\left(\alpha_i+\frac{r_i}{\sqrt{\sum_{j=1}^N\frac{r_j^2}{\alpha_j}}}\right)\qquad i=1,...,N; \hspace{0.5cm}r_1^2+...+r_N^2=1 \, ,
\end{equation}
where the unit vector $\vec{r}$ is to be considered as anchored at the centre of the ellipsoid. 
To ensure holomorphy of the superpotential we need $n_i\geq 0$, and the masses of the imaginary field components $y_i$ are given by
\begin{equation}
    m_{y_i}^2 \; = \; \frac{2^{2-3(\sum\alpha_i)}}{\alpha_i^2} \cdot \left(\alpha_i^2-\frac{r_i^2}{\left(\sum_{j=1}^{N}\frac{r_j^2}{\alpha_j}\right)}\right)
    \cdot a^2 \cdot \prod_{j=1}^Nx_j^{2n_{j}-3\alpha_j},\qquad i=1,...,N \, .
\end{equation}
For stability, we impose conditions similar to (\ref{2stability}), namely:
\begin{equation} \label{Nstability}
    \alpha_i^2-\frac{r_i^2}{\left(\sum_{j=1}^2\frac{r_j^2}{\alpha_j}\right)} \; \geq \; 0 \qquad i=1,...,N \, .
\end{equation}
As in two-field models, ensuring these stability conditions are satisfied implies that the holomorphy conditions are also satisfied. 
For a given unit vector $\vec{r}$, one can ask what values of $\alpha_1,...,\alpha_N$ satisfy the stability conditions. 
We find a multidimensional region analogous to that in Fig.~\ref{f1}, with a vertex that satisfies $\alpha_1+...+\alpha_N=1$.
We show In Fig.~\ref{q44}  the allowed region of $\alpha_1,\alpha_2$ and $\alpha_3$ for a three-field model with $\vec{r}=(1/\sqrt{3},1/\sqrt{3},1/\sqrt{3})$. 

The vertex is a special solution that corresponds to $V=0$ in both the real and imaginary field directions. 
When the sign of one of the components of $\vec{r}$ is changed, the region in $\alpha_1,...,\alpha_N$ 
space that satisfies (\ref{Nstability}) remains the same. Therefore, there are $2^N$ unit vectors, each 
differing only in the sign of the components, that have the same vertex solution.

The above observations on the vertex solution can be summarized as follows. When
\begin{equation}
    \sum_{i=1}^{N}\alpha_i \; = \; 1 \, ,
\end{equation}
there are $2^N$ superpotentials of the form
\begin{equation}
    W \; = \; a \, \phi_{i_1}^{3\alpha_{i_1}}...\phi_{i_n}^{3\alpha_{i_n}} \, ,
\end{equation}
where $\{i_1,...,i_n\}$ ($n\leq N$) is a subset of $\{1,2,...,N\}$, that all give
$V=0$ in both the real and imaginary field directions.

\subsection{De Sitter Solutions}

\begin{figure}[ht]
\centering
\includegraphics[scale=0.7]{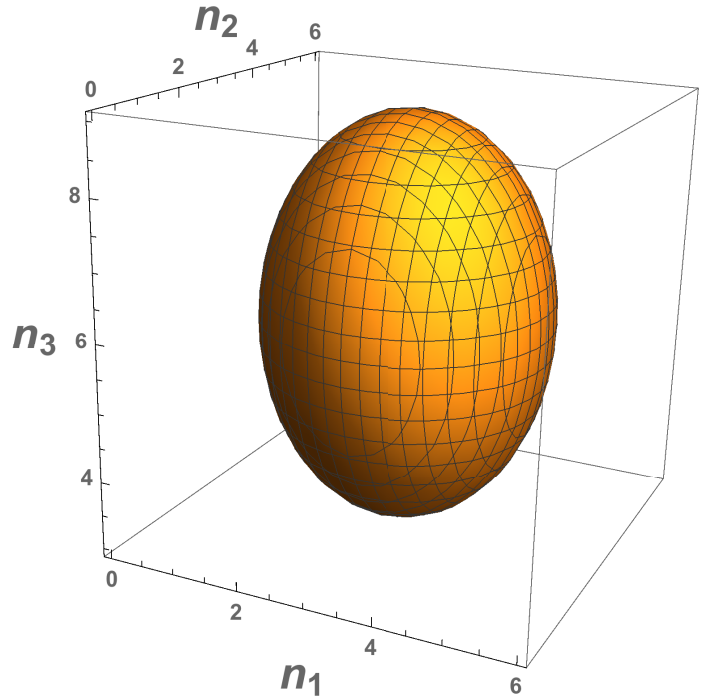}
\caption{\it Minkowski solutions for the three-field model with $\alpha_1=2$, $\alpha_2=2$ and $\alpha_3=4$.}
\label{zinc}
\end{figure}

Finally we discuss de Sitter solutions in $N$-field models. Here the K\"ahler potential is again given by
\begin{equation}
    K \; = \; - \, 3 \, \sum_{i=1}^{N} \, \alpha_i\text{ln}(\phi_i+\phi^{\dagger}_i) \, ,
\end{equation}
and, as in the two-field case, the superpotential may be constructed from two antipodal points of the ellipse (\ref{Nellipse}):
\begin{equation}
    W \; = \; a \, \left(\prod_{i=1}^{N}\phi_i^{n_{i+}}-\prod_{i=1}^{N}\phi_i^{n_{i-}} \right) \, ,
\end{equation}
where the exponents are given by
\begin{equation}
    n_{i\pm} \; = \; \frac{3}{2}\left(\alpha_i\pm\frac{r_i}{\sqrt{\sum_{j=1}^N\frac{r_j^2}{\alpha_j}}}\right)\qquad i=1,...,N; \hspace{0.5cm}r_1^2+...+r_N^2=1 \, ,
\end{equation}
and the potential along the real field directions is then
\begin{equation}
    V \; = \; 3 \cdot 2^{(2-3\sum_{i=1}^N\alpha_i)} \cdot a^2 \, .
\end{equation}
We use a simple three-field model with $\alpha_1=2$, $\alpha_2=2$ and $\alpha_3=4$ for illustration. 
The Minkowski solutions are described by an ellipsoid in $(n_1,n_2,n_3)$ space centred at $(3,3,6)$, which is shown in Fig.~\ref{zinc}.

To construct de Sitter solutions for this model, we choose the antipodal points 
$(3,3,9),(3,3,3)$ corresponding to the unit vector $\vec{r}=(0,0,1)$, which yield
the superpotential:
\begin{equation}
    W \; = \; a \, (\phi_1^{3}\phi_2^3\phi_3^9-\phi_1^{3}\phi_2^{3}\phi_3^{3}) \, .
\end{equation}
This yields a de Sitter potential along the real field directions with potential
\begin{equation}
    V \; = \; 3 \cdot 2^{-16} \cdot a^2 \, .
\end{equation}

\subsection{Stability Analysis}

The stability analysis of the de Sitter solution in the $N$-field model is difficult,
as it requires finding the eigenvalues of an $N\times N$ matrix. However, as in the two-field model, 
we do not expect the solution to be stable unless the K\"ahler potential is deformed, e.g., to
\begin{equation}
    K \; = \; - \, 3 \, \sum_{i=1}^{N} \, \alpha_i\text{ln}(\phi_i+\phi^{\dagger}_i+ b_i (\phi_i-\phi_i^{\dagger})^4) \, .
\end{equation}
With this modification, for any given unit vector $\vec{r}$ there should exist a region in $(\alpha_1,...,\alpha_N)$ 
space where the de Sitter solution is stable.

To demonstrate this in a specific three-field example, we consider the model with three chiral fields $S,T,U$ that was considered in~\cite{STU}. 
This model is defined by the following K\"ahler potential and superpotential:
\begin{equation}
\begin{split}
    K&= \; - \, \text{ln}(S+S^{\dagger})\, - \, 3\text{ln}(T+T^{\dagger}) \, - \, 3\text{ln}(U+U^{\dagger}) \, ,\\
    W&= \; W(S,T,U) \, .
\end{split}    
\end{equation}
This model is of particular interest as it arises in the compactification of Type IIB string theory on $T^6/\mathbb{Z}_2\times \mathbb{Z}_2$. 
Then the three chiral fields are the axiodilaton $S$, a volume modulus $T$ and a complex structure modulus $U$. 
One expects that the perturbative contribution to the superpotential should be a polynomial and that
the non-perturbative contribution would have a decaying exponential form. For our analysis we assume that the 
powers of the fields in the superpotential could also be fractional. In our notations, this $STU$ model has 
$\alpha_1=1/3, \alpha_2=1$ and $\alpha_3=1$.

We first construct a Minkowski solution. We can use the stability conditions to find a unit vector $\vec{r}$ and 
construct an appropriate superpotential. One such unit vector is $\vec{r}=(0,1,0)$. This leads to a superpotential of the form
\begin{equation}
    W \; = \; a S^{1/2}T^{3}U^{3/2} \, ,
\end{equation}
which gives a stable Minkowski solution $V=0$ along real field directions~\footnote{We mention
in passing that the $STU$ model does not admit any superpotential with only integer powers,
for either Minkowski or de Sitter solutions.}.

In order to construct de Sitter solutions we add stabilization terms to the K\"ahler potential:
\begin{equation}
    K \; = \; - \, \text{ln}(S+S^{\dagger}+ b_S (S-S^{\dagger})^4) \, - \, 3 \, \text{ln}(T+T^{\dagger}+ b_T (T-T^{\dagger})^4) \, - \, 3\, \text{ln}(U+U^{\dagger}+ b_U (U-U^{\dagger})^4) \, .
\end{equation}
As discussed above, we use antipodal points to construct the superpotential, choosing $\vec{r}=(0,\pm1,0)$, in which case:
\begin{equation}
    W \; = \; a \, S^{1/2}U^{3/2}(T^3-1) \, .
\end{equation}
With this we get a de Sitter solution along the real field directions with potential
\begin{equation}
    V \; = \; \frac{3}{32} \cdot a^2 \, .
\end{equation}
In order to check whether the de Sitter solution is stable for the antipodal points that we have chosen, 
we calculate the Hessian matrix along the real field directions to verify that the eigenvalues are non-negative. Defining
\begin{equation}
    S \; = \; s+iy_1 \, , \qquad T \; = \; t+iy_2 \, ,\qquad U\; = \; u+iy_3 \, ,
\end{equation}
we calculate the Hessian matrix $\partial^2 V / \partial y_i \partial y_j: i, j = 1, 2, 3$ along the real field directions, finding
\begin{equation}
\begin{bmatrix}
   \frac{a^2(1+4t^3+t^6)}{64s^2t^3} & 0 &
    0\\
    0 & \frac{-3a^2+72a^2b_T(1+t^3)^2}{16t^2} &
    0\\
    0 &
    0 &
    \frac{3a^2(1+4t^3+t^6)}{64t^3u^2}
\end{bmatrix} \, .
\end{equation}
We see that the Hessian matrix is diagonal, so the eigenvalues are simply the diagonal entries. 
For the Hessian matrix to be positive semi-definite we need
    \begin{equation}
        - \, 3a^2+72 a^2 \, b_T \, (1+t^3)^2 \; \geq \; 0 \, ,
 \end{equation}
which is independent of $b_S$ and $b_U$. Therefore, we simply need
    \begin{equation}
        b_T \; \geq \; \frac{1}{24} \, ,
\end{equation}
with no restriction on $b_S$ and $b_U$.

\section{Conclusion and Outlook} \label{sec:conx}

Generalizing previous discussions of de Sitter solutions in single-field no-scale models~\cite{EKN1,rs,eno9},
in this paper we have discussed de Sitter solutions in multi-field no-scale models as may appear in realistic
string compactifications with multiple moduli. 

As a preliminary,
we shoed that the space of Minkowski vacua in multi-field no-scale models is characterized by the surface of an ellipsoid. 
The parameters in these models are the coefficients $(\alpha_1,...,\alpha_N)$ in the generalized no-scale
K\"ahler potential and a unit vector $\vec{r}$ that selects a particular pair of antipodal points on this ellispoid whose center
is located at  $(3\alpha_1/2,...,3\alpha_N/2)$. 
Requiring the stability of Minkowski solutions for a fixed $\vec{r}$ leads us to a 
region in $(\alpha_1,...,\alpha_N)$ space with a vertex that is a special point where $\sum_{i=1}^{N}\alpha_i=1$.
Such points describe Minkowski vacua with potentials that are flat in both the real and imaginary field directions. 
In this way we constructed $2^N$ monomial (in each field) superpotentials 
for models with $\sum_{i=1}^{N}\alpha_i=1$ that yield acceptable Minkowski vacua. The exponent of each monomial
is determined by the coefficients $\alpha_i$ and the vectors, $r_i$. 

We then constructed de Sitter solutions by combining the superpotentials at antipodal points,
generalizing a construction given originally in the single-field case in~\cite{EKN1}. 
These de Sitter solutions are unstable if the simple no-scale K\"ahler potential is used, and require stabilization. 
We showed that modifying the K\"ahler potential with a quartic term stabilizes a specific two-field model with 
$\alpha_1=\alpha_2=\alpha$ and $\vec{r}=(1/\sqrt{2},1/\sqrt{2})$ for $\alpha\geq2/3$, and
we expect the stability to hold for other generic $\vec{r}$ for suitable ranges of $\alpha_1,\alpha_2$.
We also expect that similar stable de Sitter solutions exist for $N$-field models under certain conditions,
as demonstrated explicitly in a specific three-field model motivated by the compactification of Type IIB string theory~\cite{STU}.

We note that satisfying the stability requirement also ensures that the superpotential is holomorphic in the Minkowski case, i.e.,
contains only positive powers of the chiral fields, whereas this is not necessarily true in the de Sitter case. It is easy to find infinite discrete series of models for
which these powers are integral, and we have provided a number of illustrative single- and multi-field examples.

As noted in the Introduction, it is currently debated whether string theory admits de Sitter solutions~\cite{string}.
If this were not the case, measurements of the accelerating expansion of the Universe~\cite{cc} and the
continuing success of cosmological inflation~\cite{inf} would suggest that our Universe lies in the swampland. Our
working hypothesis is that this is not the case, and that deeper understanding of string theory will reveal
how it can accommodate de Sitter solutions. Since no-scale supergravity is the appropriate framework
for discussing cosmology at scales hierarchically smaller than the string scale, assuming also that $N = 1$ supersymmetry
holds down to energies $\ll m_{\rm Planck}$, the explorations in this paper may provide a helpful guide
to the structure of the low-energy effective field theories of de Sitter string solutions. As such, they
may even provide some useful signposts towards the construction of such solutions.

\section*{Acknowledgements}

B.N. thanks William Linch III and Daniel Butter for useful discussions.
The work of J.E. was supported in part by STFC (UK) via the research grant ST/L000258/1
and in part by the Estonian Research Council via a Mobilitas Pluss grant. 
The work of B.N. was supported by the Mitchell/Heep Chair in High Energy Physics, Texas A\&M University.
The work of D.V.N. was supported in part by the DOE
grant DE-FG02-13ER42020 and in part by the Alexander~S.~Onassis Public
Benefit Foundation. The work of K.A.O. was supported in part by
DOE grant DE-SC0011842 at the University of Minnesota.

\end{document}